\documentclass{article}
\usepackage{spconf,amsmath,graphicx,hyperref}
\usepackage{multirow}
\usepackage[table,xcdraw]{xcolor}
\usepackage{booktabs}
\usepackage{siunitx}  % 用于数字对齐
\usepackage{enumitem}
\usepackage{amssymb}
\usepackage{ifsym}
\usepackage{mathrsfs} % 额外花体字体包

\usepackage{inconsolata}
\usepackage[flushmargin]{footmisc}
\makeatletter
\def\thanks#1{\protected@xdef\@thanks{\@thanks\protect\footnotetext{#1}}}
\makeatother

\title{Attention-weighted Centered Kernel Alignment for Knowledge Distillation in Large Audio-Language Models Applied to Speech Emotion Recognition}
\name{
  \begin{tabular}{c}
    Qingran Yang$^{1,2\dagger}$, Botao Zhao$^{1\dagger}$, Zuheng Kang$^1$, Xue Li$^2$, Yayun He$^1$, Chuhang Liu$^1$, \\
    Xulong Zhang$^1$,  Xiaoyang Qu$^1$, Junqing Peng$^1$, Jianzong Wang$^{1\ast}$
  \end{tabular}
  \vspace{-0.5cm}
\thanks{This work is supported by Shenzhen-Hong Kong Joint Funding Project (Category A) under grant No. SGDX20240115103359001. \\
 $^\dagger$ These authors contributed equally to this work. \\
 $^\ast$ Corresponding author: Jianzong Wang (jzwang@188.com).}
}

\address{
$^1$Ping An Technology (Shenzhen) Co., Ltd., Shenzhen, China,\\
$^2$Harbin Institute of Technology, Harbin, China.
}
\begin{document}
\ninept
\maketitle
\begin{abstract}
The emergence of Large Audio-Language Models (LALMs) has advanced Speech Emotion Recognition (SER), but their size limits deployment in resource-constrained environments. While Knowledge Distillation is effective for LALM compression, existing methods remain underexplored in distilling the cross-modal projection module (Projector), and often struggle with alignment due to differences in feature dimensions. We propose PL-Distill, a KD framework that combines Projector-Level Distillation (PDist) to align audio embeddings and Logits-Level Distillation (LDist) to align output logits. PDist introduces Attention-weighted Centered Kernel Alignment, a novel approach we propose to highlight important time steps and address dimension mismatches. Meanwhile, LDist minimizes the Kullback-Leibler divergence between teacher and student logits from audio and text modalities. On IEMOCAP, RAVDESS, and SAVEE, PL-Distill compresses an 8.4B-parameter teacher to a compact 1.1B-parameter student, consistently outperforming the teacher, state-of-the-art pretrained models, and other KD baselines across all metrics.
\end{abstract}
\section{Introduction}
\label{sec:intro}

Speech Emotion Recognition (SER) technology aims to detect emotions in speech, which shows promise in mental health, customer service, and medical fields \cite{c1,SER-1,SER-3}, attracting extensive research and attention.
Conventional SER models typically leverage classic deep learning architectures, such as CNN, RNN, and LSTM. With the advent of self-supervised learning, models like WavLM \cite{c3} and Whisper \cite{c4} enhance SER by learning generalizable audio representations.
While effective for learning acoustic features, these methods remain limited in their ability to capture the textual semantics embedded in speech. This gap restricts their capacity to interpret complex emotions, which often depend heavily on linguistic content rather than just acoustic cues.
The development of Large Audio-Language Models (LALMs), such as Qwen2-Audio \cite{c5,EMO-RL}, presents a new avenue for addressing this limitation in SER.
By integrating Large Language Models (LLMs) to fuse acoustic features with semantic information, these models enable a deeper understanding of complex emotions \cite{C7}. Despite these advances, the large parameter scale of LALMs, such as Qwen2-Audio with 8.4B parameters \cite{c5}, leads to high inference costs, restricting their applicability in resource-constrained scenarios.

Knowledge Distillation (KD) \cite{c8,KD-1} is a model compression technique that can reduce the large parameter scale of LALMs by transferring knowledge from powerful teacher models to lightweight student models, enabling the student to match the teacher’s performance with fewer parameters and lower computation.
Most existing KD methods focus solely on the text modality, using Forward Kullback-Leibler Divergence (Forward KL) to align teacher-student output logits by minimizing KL divergence. Reverse KL \cite{c9} and Jensen-Shannon Divergence \cite{c10} refine this alignment by improving distribution measurement.
Recent advances have extended KD to Multimodal Large Language Models (MLLMs), as seen in LLaVA-MoD \cite{C12} and LLAVA-KD \cite{llava-kd}. LLaVA-MoD optimizes distillation via the MoE \cite{moe} architecture but increases structural complexity and resource use.
In contrast, LLAVA-KD adjusts distillation strategies without modifying model structure, effectively leveraging multimodal information.

However, despite LALMs being a critical type of MLLMs, research on their distillation has been overlooked. Furthermore, directly applying distillation methods for MLLMs to LALMs presents several challenges as follows: (1) Most are designed for visual tasks, emphasizing spatial dependencies (e.g., inter-object interactions) \cite{llava-kd}, and cannot fully capture the temporal importance for audio. (2) The Projector, crucial for cross-modal fusion by mapping non-text features into language space, is often excluded from distillation. Even when Projector features are distilled, dimensional mismatches between teacher and student outputs remain a challenge \cite{c15}.

To address these issues, we propose PL-Distill, a lightweight knowledge distillation framework for LALMs in SER. PL-Distill consists of two main components: (1) Projector-Level Distillation (PDist): We introduce Attention-weighted Centered Kernel Alignment (AwCKA), an extension of CKA \cite{c16}, which incorporates dynamic weights based on the teacher model's self-attention scores. The proposed AwCKA not only addresses feature dimension mismatches but also prioritizes emotionally significant time steps. (2) Logits-Level Distillation (LDist): This component minimizes the KL divergence between the teacher and student logits from both audio and text modalities, ensuring cross-modal consistency while facilitating the transfer of both acoustic and textual knowledge. We summarize our contributions as follows:

\begin{itemize}[left=0pt]
    \item We introduce PL-Distill, a novel framework filling the gap in LALM knowledge distillation research and significantly reducing deployment costs in resource-constrained settings.
    \item We propose AwCKA to align the audio embeddings from teacher and student, which helps the projector focus on critical audio tokens, and solves the feature dimension mismatches.
    \item Extensive experiments demonstrate the proposed methods could compress the Qwen2-Audio teacher model (8.4 billion parameters) by 87\% and achieve better performance than current state-of-the-art models on authoritative SER datasets.
\end{itemize}

\begin{figure*}[t!] % 使用 figure* 占两列，[t!] 强制置顶
  \centering % 图片居中
  % \label{f1dwqd}
  \includegraphics[width=\textwidth]{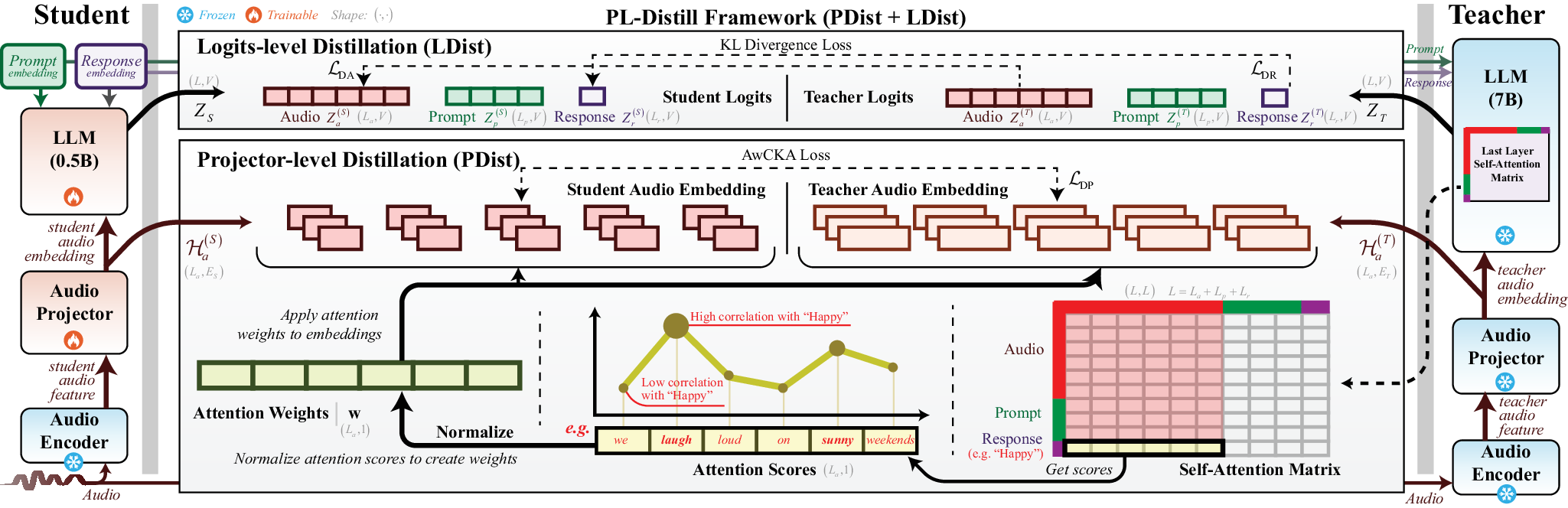} % 替换为你的图片文
  \vspace{-0.6cm}
  \caption{An overview of our PL-Distill framework, which includes Projector-Level Distillation (PDist) and Logits-Level Distillation (LDist).} % 图片标题
  \label{f1dwqd}
  \vspace{-0.4cm}
\end{figure*}

\vspace{-0.3cm}
\section{Methodology}
\label{sec:format}
As shown in Figure \ref{f1dwqd}, our PL-Distill includes Projector-Level and Logits-Level Distillation. We introduce our method via the student model architecture, and these two distillation components.

\vspace{-0.3cm}
\subsection{Architecture of the Student Model}
\label{ssec:model_architecture}
We use Qwen2-Audio (8.4B parameters) as the teacher model, and the student model follows the same architecture but has only 1.1B parameters, about 13\% of the teacher’s size. The specific structure of the models is as follows:
\begin{itemize}[left=0.2pt]
    \item \textbf{Audio Encoder}: Both models utilize the Whisper large v3 \cite{c4} model as the audio encoder. During training, we load the pretrained parameters from the teacher model and freeze this module.
    \item \textbf{Audio Projector}: Both models employ a single-layer MLP as the Audio Projector, which maps the audio features from the audio encoder to their respective LLM embedding spaces, producing the audio embeddings.
    \item \textbf{Large Language Model (LLM)}: The teacher model uses Qwen2-7B, while the student uses the smaller Qwen2-0.5B, initialized from Qwen2-0.5B-Instruct to leverage its language capabilities.
\end{itemize}

For multimodal integration, the audio embeddings from the Audio Projector are concatenated with text embeddings to form a joint representation \( H = [\mathcal{H}_a, \mathcal{H}_p, \mathcal{H}_r] \), where \( \mathcal{H}_a \), \( \mathcal{H}_p \) and \( \mathcal{H}_r \) represent the audio, prompt and response embeddings \cite{prompt}. This representation is processed by the LLM to produce output logits \( Z = [Z_a, Z_p, Z_r] \), corresponding to the audio, prompt, and response.

\subsection{Projector-Level Distillation}
\label{ssec:Projector_dis}
For an audio sample, both the teacher and student models use the same Audio Encoder to extract audio features. These features are then processed by their respective Audio Projectors to obtain the teacher's audio embedding \(\mathcal{H}_{a}^{(T)} \in \mathbb{R}^{L_a\times E_T}\) and the student's audio embedding \(\mathcal{H}_{a}^{(S)} \in \mathbb{R}^{L_a\times E_S}\). Here, \(L_a\) denotes the length of the audio token sequence, \(E_T\) and \(E_S\) represent the dimensions of the audio tokens, corresponding to the teacher's and student's respective LLM embedding dimensions.

In Projector-Level Distillation (PDist), to distill \(\mathcal{H}_{a}^{(T)}\) and \(\mathcal{H}_{a}^{(S)}\) despite their differing dimensions (\(E_T \neq E_S\)), we employ Centered Kernel Alignment (CKA) \cite{c16}.
CKA is a similarity metric that measures representation alignment through covariance structures, making it particularly effective for comparing representations with different dimensionalities as it focuses on statistical dependencies rather than raw feature values \cite{cka}. We employ a linear kernel implementation of CKA because previous work \cite{c16,c17} has not observed significant accuracy improvements with non-linear kernels, while linear kernels offer substantially greater computational efficiency, which is particularly valuable for scaling to large models.

While standard linear CKA uniformly treats all audio tokens, for tasks like SER, emotional cues are often concentrated in specific key time steps within the audio token sequence. To account for this temporal sparsity and focus distillation on more salient parts of the audio, we introduce importance weights for each audio token, leading to Attention-weighted Centered Kernel Alignment (AwCKA).

The audio token weights, denoted as \( \mathbf{w} \in \mathbb{R}^{L_a \times 1} \), are derived from the final self-attention layer of the teacher model's LLM. The choice of the last layer is made because it reflects the comprehensive and global semantic information and provides a more holistic understanding relevant to the task \cite{c18}. We extract the attention scores from the response token (e.g., 'Happy') to the audio tokens as the raw values for \( \mathbf{w} \). 
These attention scores, based on the transformer's positional encoding and self-attention mechanism, inherently capture long-term temporal dependencies. 
A higher attention score indicates greater relevance to the response and a more critical role for the corresponding audio token in conveying the emotion. Finally, we normalize these attention scores to obtain the audio token weights \( \mathbf{w} \), as follows:
\[
\mathbf{w} = \frac{\mathbf{A}}{\sum_{i=1}^{L_a} A_i},
\tag{1}
\]
where \( \mathbf{A} \in \mathbb{R}^{L_a \times 1} \) represents the attention scores, and \( \sum_{i=1}^{L_a} A_i \)
is the sum of all attention scores.

Then, we apply the audio token weights $\mathbf{w}$  to the audio embeddings $\mathcal{H}_{a}^{(T)}$ and $\mathcal{H}_{a}^{(S)}$, yielding the weighted embeddings for both teacher and student: ${H}_T \in \mathbb{R}^{L_a \times E_T}$ and ${H}_S \in \mathbb{R}^{L_a \times E_S}$.
\[
{H}_T = \mathbf{w} \odot \mathcal{H}_{a}^{(T)}, \quad{H}_S = \mathbf{w} \odot \mathcal{H}_{a}^{(S)},
\tag{2}
\]
where \(\odot\) denotes multiplication with broadcasting along the embedding dimension, each scalar in \(\mathbf{w} \in \mathbb{R}^{L_a \times 1}\) scaling all embedding dimensions of the corresponding audio token in \(\mathcal{H}_{a}^{(T)}\) and \(\mathcal{H}_{a}^{(S)}\).

The first step in computing linear CKA is to center the weighted embeddings by subtracting the mean of each feature along the token sequence dimension. This step results in \( \hat{H}_T \) and \( \hat{H}_S \), which correspond to the centered embeddings for the teacher and student models, respectively. After obtaining the centered embeddings \( \hat{H}_T \) and \( \hat{H}_S \), we use the Hilbert-Schmidt Independence Criterion (HSIC) \cite{HSIC} to measure the similarity between two features by comparing their covariance.
Then, the AwCKA between the audio embeddings of the teacher and the students is defined as:
\begin{align*}
&\text{AwCKA}(\mathcal{H}_{a}^{(T)},\, \mathcal{H}_{a}^{(S)},\, \mathbf{w})
= \text{CKA}({H}_T, {H}_S) \\ % AwCKA 定义为对加权嵌入应用的 CKA
&= \frac{\text{HSIC}({H}_T, {H}_S)}{\sqrt{\text{HSIC}({H}_T, {H}_T)} \cdot \sqrt{\text{HSIC}({H}_S, {H}_S)}} \\ % CKA 定义为归一化的 HSIC
&= \frac{\|\hat{H}_T^T \hat{H}_S \|_F^2}{\|\hat{H}_T^T \hat{H}_T \|_F \cdot \|\hat{H}_S^T \hat{H}_S \|_F},
\tag{3}
\end{align*}
where \( \|\cdot\|_F \) denotes the Frobenius norm. It can be shown that $\text{AwCKA}(\mathcal{H}_{a}^{(T)},\, \mathcal{H}_{a}^{(S)},\, \mathbf{w}) \in [0, 1]$ \cite{c16}, with higher values indicating greater similarity between the two feature representations.

Finally, the distillation loss for the Audio Projector outputs is defined as:
\begin{align*}
\mathcal{L}_{\text{DP}}
&= 1 - \text{AwCKA}(\mathcal{H}_{a}^{(T)},\, \mathcal{H}_{a}^{(S)},\, \mathbf{w}).
\tag{4}
\end{align*}
% where \( \|\cdot\|_F \) denotes the Frobenius norm. It can be shown that $\text{AwCKA}(\mathcal{H}_{a}^{(T)},\, \mathcal{H}_{a}^{(S)},\, \mathbf{w}) \in [0, 1]$ \cite{c16}, with higher values indicating greater similarity between the two feature representations.
% Finally, The distillation loss for the Audio Projector outputs is defined as:
% \begin{align*}
% \mathcal{L}_{\text{DP}}
% &= 1 - \text{AwCKA}(\mathcal{H}_{a}^{(T)},\, \mathcal{H}_{a}^{(S)},\, \mathbf{w})
% \tag{4}
% \end{align*}

\subsection{Logits-Level Distillation}
\label{ssec:logits_dis}
As described in Section~\ref{ssec:model_architecture}, for an audio sample, the LLM outputs logits
\( Z = [\,Z_a \in \mathbb{R}^{L_a \times V},\; Z_p \in \mathbb{R}^{L_p \times V},\; Z_r \in \mathbb{R}^{L_r \times V}\,] \),
where \(L_a\), \(L_p\), and \(L_r\) denote the sequence lengths of the audio, prompt, and response tokens, \(V\) is the vocabulary size, and the total sequence length is \(L = L_a + L_p + L_r\). We denote \( Z_S\) \ and \( Z_T \) as the logits for the student and teacher models. It is worth noting that the logits from both the teacher and student models have the same sequence length and vocabulary size.

In Logits-Level Distillation (LDist), following recent studies on LLM distillation, we exclude prompt tokens from the distillation loss, as the student model is not required to predict them. 
For the audio logits from the teacher and student models, denoted as \(Z_a^{(T)}\) and \(Z_a^{(S)}\), respectively,
we employ temperature-scaled softmax to obtain the corresponding probability distributions \(P_a^{(T)}\) and \(P_a^{(S)}\).
We then compute the Kullback-Leibler (KL) divergence between these distributions as the distillation objective, defining the loss as \(\mathcal{L}_{\text{DA}}\).
The calculation process is as follows:
\[
\mathcal{L}_{\text{DA}} = \text{KL}(P_a^{(T)} \parallel P_a^{(S)}) = \sum_{k=1}^{V} P_a^{(T)}(k) \log \left( \frac{P_a^{(T)}(k)}{P_a^{(S)}(k)} \right),
\]
with 
\[
P_a(k) = \frac{\exp(Z_a(k) / t)}{\sum_{j=1}^{V} \exp(Z_a(j) / t)}.
\tag{5}
\]
Here, \( V \) is the vocabulary size for the LLM, while \( Z_a(k) \) and \( P_a(k) \) represent the logit and probability values for the \( k \)-th token in the vocabulary, respectively. The temperature \( t \) controls the smoothness of the distribution: a higher \( t \) leads to a softer distribution that reveals more information, while a lower \( t \) produces a sharper distribution that emphasizes the most probable tokens \cite{c8}.

Similar to the audio logits, for the response logits, we define \( Z_r^{(T)} \) for the teacher model and \( Z_r^{(S)} \) for the student model, with \( P_r^{(T)} \) and \( P_r^{(S)} \) representing their corresponding probability distributions. The loss function, denoted as \( \mathcal{L}_{\text{DR}} \), can be expressed as:
\[
\mathcal{L}_{\text{DR}} = \text{KL}(P_r^{(T)} \parallel P_r^{(S)}) = \sum_{k=1}^{V} P_r^{(T)}(k) \log \left( \frac{P_r^{(T)}(k)}{P_r^{(S)}(k)} \right).
\tag{6}
\]

In knowledge distillation, the student model not only distills knowledge from the teacher model but also learns from the true emotion labels \cite{c8}. The cross-entropy loss, denoted as \( \mathcal{L}_{\text{CE}} \), is used to measure the difference between the probability distributions of the student response logits and the ground truth labels. The \( \mathcal{L}_{\text{CE}} \) is defined as:
\[
\mathcal{L}_{\text{CE}} = - \sum_{i=1}^{L_r} \sum_{k=1}^{V} y_i(k) \log \left( P_r^{(S)}(i,k) \right).
\tag{7}
\]
Here, \( L_r \) is the length of the response token sequence. \( P_r^{(S)}(i,k) \) is the probability of the \( i \)-th response token being the \( k \)-th token in the vocabulary, as predicted by the student model. \( y_i(k) \) is the indicator for the true label of the \( i \)-th token, where \( y_i(k) = 1 \) if the \( i \)-th true label token corresponds to the \( k \)-th token in the vocabulary, and \( y_i(k) = 0 \) otherwise.

The total objective function is:
\[
\mathcal{L}_{\text{total}} =  \mathcal{L}_{\text{CE}} + \alpha \mathcal{L}_{\text{DP}} + \beta \mathcal{L}_{\text{DA}} + \gamma \mathcal{L}_{\text{DR}},
\tag{8}
\]
where \( \mathcal{L}_{\text{DP}} \) is from PDist, \( \mathcal{L}_{\text{DA}} \) and \( \mathcal{L}_{\text{DR}} \) are from LDist, and \( \mathcal{L}_{\text{CE}} \) is from the true emotion labels.

\section{Experimental Results}
\label{sec:Experimental Results}

\begin{table*}[t] % [t] 尝试将表格放置在页面顶部
  \centering % 居中表格
  \fontsize{8}{9}\selectfont % 设置字体大小和行距
  \caption{Comparison of main performance metrics for various models on the IEMOCAP, RAVDESS, and SAVEE datasets. Results for pretrained models are cited from the Emobox benchmark \cite{c20}. The baseline for comparison refers to the Forward KL method. SOTA results among Pretrained Models are highlighted in \textbf{bold}, and the overall best results across all models are highlighted in \textbf{\textcolor[HTML]{CB0000}{red}}.} % 请替换为您的表格标题
  \label{t1} % 请替换为您的表格标签
\begin{tabular}{@{}ccccccccccc@{}}
    \toprule
                                         &                                     & \multicolumn{3}{c}{\textbf{IEMOCAP}} & \multicolumn{3}{c}{\textbf{RAVDESS}} & \multicolumn{3}{c}{\textbf{SAVEE}} \\
    \cmidrule(lr){3-5} \cmidrule(lr){6-8} \cmidrule(lr){9-11} % 细化分栏线条
    \multirow{-2}{*}{\textbf{Type}}          & \multirow{-2}{*}{\textbf{Model}}    & \textbf{UA(\%)}                           & \textbf{WA(\%)}                           & \textbf{F1(\%)}                           & \textbf{UA(\%)}                           & \textbf{WA(\%)}                           & \textbf{F1(\%)}                           & \textbf{UA(\%)}                           & \textbf{WA(\%)}                           & \textbf{F1(\%)}                           \\ \midrule
                                         & WavLM large \cite{c3}                         & 69.47                                                         & 69.07                                                         & 69.29                                                         & 72.00                                                         & 72.22                                                         & 71.42                                                         & 75.65                                                         & 78.25                                                         & {\textbf{78.38}}                                  \\
                                         & data2vec 2.0 large \cite{c19}                  & 57.30                                                         & 56.23                                                         & 56.70                                                         & 71.15                                                         & 71.63                                                         & 70.94                                                         & {\textbf {75.75}}                                  & {\textbf {78.59}}                                  & 78.24                                                         \\
    \multirow{-3}{*}{Pretrained Models} & Whisper large v3 \cite{c4}                  & {\textbf {73.54}}                                  & {\textbf {72.86}}                                  & {\textbf {73.11}}                                  & {\textbf {75.32}}                                  & {\textbf {75.87}}                                  & {\textbf {75.19}}                                  & 74.07                                                         & 77.24                                                         & 75.31                                                         \\ \midrule
    Teacher Model                            & Qwen2-Audio \cite{c5}                         &  64.33                                                        & 60.37                                                         & 61.61                                                        & 63.67                                                         & 63.33                                                         & 60.84                                                         & 61.43                                                         & 58.33                                                         & 53.92                                                         \\ \midrule
                                         & SFT                                 & 77.59                                                         & 76.04                                                         & 76.29                                                         & 82.67                                                         & 76.67                                                         & 74.48                                                         & 78.10                                                         & 80.83                                                         & 76.70                                                         \\
                                         & Forward KL                                  & 79.33                                                         & 76.77                                                         & 77.22                                                         & 85.54                                                         & 87.08                                                         & 84.60                                                         & 83.70                                                         & 85.00                                                         & 83.07                                                         \\
                                         & Reverse KL \cite{c9}                                 & 79.32                                                         & 78.34                                                         & 78.76                                                         & 87.74                                                         & 86.67                                                         & 84.03                                                         & 79.05                                                         & 81.67                                                         & 77.70                                                         \\
                                         & LLaVA-KD \cite{llava-kd}                            & 81.56                                                         & 80.28                                                         & 80.49                                                         & 89.36                                                         & 88.75                                                         & 88.03                                                         & 85.71                                                         & 87.50                                                         & 85.55                                                         \\
    \multirow{-5}{*}{Student Models}             & \textbf{PL-Distill (Ours)}                & {\color[HTML]{CB0000} \textbf{83.91}} & {\color[HTML]{CB0000} \textbf{82.12}} & {\color[HTML]{CB0000} \textbf{82.62}} & {\color[HTML]{CB0000} \textbf{92.58}} & {\color[HTML]{CB0000} \textbf{92.08}} & {\color[HTML]{CB0000} \textbf{92.23}} & {\color[HTML]{CB0000} \textbf{91.43}} & {\color[HTML]{CB0000} \textbf{92.50}} & {\color[HTML]{CB0000} \textbf{91.36}} \\
    \midrule % 在新数据之前添加一条分隔线
                                         & Ours VS  SOTA                    & {\color[RGB]{0,127,127} $\uparrow$14.1\%}        & {\color[RGB]{0,127,127} $\uparrow$12.7\%}        & {\color[RGB]{0,127,127} $\uparrow$13.0\%}        & {\color[RGB]{0,127,127} $\uparrow$22.9\%}        & {\color[RGB]{0,127,127} $\uparrow$21.4\%}        & {\color[RGB]{0,127,127} $\uparrow$22.7\%}        & {\color[RGB]{0,127,127} $\uparrow$20.7\%}        & {\color[RGB]{0,127,127} $\uparrow$17.7\%}        & {\color[RGB]{0,127,127} $\uparrow$16.6\%}        \\
    \multirow{-2}{*}{Comparison}         & Ours VS Baseline                 & {\color[RGB]{0,127,127} $\uparrow$5.77\%}         & {\color[RGB]{0,127,127} $\uparrow$6.97\%}        & {\color[RGB]{0,127,127} $\uparrow$6.99\%}        & {\color[RGB]{0,127,127} $\uparrow$8.23\%}        & {\color[RGB]{0,127,127} $\uparrow$5.74\%}        & {\color[RGB]{0,127,127} $\uparrow$9.02\%}        & {\color[RGB]{0,127,127} $\uparrow$10.3\%}        & {\color[RGB]{0,127,127} $\uparrow$8.82\%}         & {\color[RGB]{0,127,127} $\uparrow$10.0\%}       \\
    \bottomrule
        \vspace{-0.6cm}
    \end{tabular}
\end{table*}

\begin{table}[t] % 使用 table 环境表示单列表格，[t] 尝试放置在页面顶部
  \centering % 居中表格
  \fontsize{8}{9}\selectfont % 设置字体大小和行距，保持与之前格式一致
  \caption{Ablation study on IEMOCAP, RAVDESS, and SAVEE datasets, evaluating LDist and PDist (with CKA or AwCKA). Performance metrics are UA and WA (e.g., IEUA is IEMOCAP UA).}% 请替换为您的表格标题
  \label{t2} % 请替换为您的表格标签
  \setlength{\tabcolsep}{3pt} % 缩小列间距，您可以尝试不同的值如 2pt, 4pt 等
\begin{tabular}{@{}c c c c c c c@{}} % 7列：1个Method列，6个数据列
    \toprule % 表格顶部线
    \textbf{Method} & \textbf{IEUA} & \textbf{IEWA} & \textbf{RAUA} & \textbf{RAWA} & \textbf{SAUA} & \textbf{SAWA} \\
    \midrule % 第二行线
    LDist                   & 80.70          & 79.72          & 88.24          & 86.25          & 83.81          & 85.83          \\
    LDist + PDist (CKA)        & 81.22          & 81.11          & 91.72          & 91.25          & 87.62          & 89.17          \\
    LDist + PDist (AwCKA)      & \textbf{83.91} & \textbf{82.12} & \textbf{92.58} & \textbf{92.08} & \textbf{91.43} & \textbf{92.50} \\
    \bottomrule % 表格底部线
\vspace{-0.7cm}
\end{tabular}
\end{table}
\subsection{Experimental Setup}
\label{ssec:Experimental Setup}
Regarding datasets, we utilize three widely adopted SER datasets for our analysis: IEMOCAP \cite{c21}, consisting of 5,531 utterances from four emotions; RAVDESS \cite{c22}, an audio-visual corpus with 4,800 speech and song samples spanning eight emotions; and SAVEE \cite{c23}, containing 480 utterances of seven emotions in both natural and exaggerated expressions.

To ensure fair and highly comparable evaluations, we strictly follow the data partitioning requirements of the Emobox Benchmark \cite{c20}, which is recognized as the most extensive and standardized SER benchmark to date.
% , comprehensively covering the performance of most state-of-the-art pretrained models.

For the training strategy, we freeze the student model's Audio Encoder. Only the student model's Audio Projector and LLM (Qwen2-0.5B) are trained, utilizing LoRA for efficient fine-tuning. Regarding training parameters, the loss function coefficients $\alpha$, $\beta$, and $\gamma$ are set to $1.0$, $0.8$, and $1.0$. The distillation temperature \( t \) is set to $2$. LoRA-specific parameters include a rank $r=8$ and $\text{lora\_alpha}=256$. We employ a batch size of $1$ with $16$ gradient accumulation steps and train for five epochs, selecting the best-performing model for final analysis.

\vspace{-0.2cm}
\subsection{Baselines and Metrics}
\label{ssec:Baselines and Metrics}
To establish baselines and assess training strategies, we compare three top-performing pretrained speech models from the Emobox Benchmark: WavLM large \cite{c3}, data2vec 2.0 large \cite{c19}, and Whisper large v3 \cite{c4}. Additionally, we introduce a teacher model (Qwen2-Audio) without additional training or fine-tuning as a reference.
In addition to our distillation framework (PL-Distill), we train student models using several strategies for comparison. These include Supervised Fine-Tuning (SFT), widely-used LLM distillation techniques such as Forward KL and Reverse KL \cite{c9}, as well as the cutting-edge MLLM distillation framework, LLaVA-KD. Model performance is evaluated with three standard metrics: Unweighted Accuracy (UA), which measures the mean of correctly classified instances; Weighted Accuracy (WA), the ratio of correct predictions to total samples; and the Macro-F1 (F1), the harmonic mean of precision and recall.

\subsection{Main Performance}
\label{ssec:Main Performance}
As shown in Table \ref{t1}, our distillation framework, PL-Distill, produces a student model with 1.1 billion parameters that significantly outperforms the teacher model (8.4 billion parameters), current state-of-the-art pretrained models, and other mainstream distillation methods across all evaluation metrics (UA, WA, F1) and datasets (IEMOCAP, RAVDESS, and SAVEE).
Compared to pretrained models, our student model shows substantial improvements across all metrics. On the RAVDESS dataset, it outperforms the best pretrained model by 22.9\%, 21.4\%, and 22.7\% in UA, WA, and F1, respectively. Even the student model fine-tuned with only SFT matches the best pretrained models. These results confirm the effectiveness of the LLM module in combining acoustic and semantic features, demonstrating the potential of LALM in speech emotion recognition.
PL-Distill also surpasses mainstream LLM distillation methods, such as Forward KL and Reverse KL. For example, on the SAVEE dataset, our method shows improvements of 10.3\%, 8.82\%, and 10.0\% across the three metrics compared to Forward KL. This is because traditional methods focus only on the text modality, neglecting non-text modalities. In contrast, our strategies, PDist and LDist, incorporate the audio modality, demonstrating its importance in improving performance.

Finally, compared with the cutting-edge MLLM distillation framework LLaVA-KD, PL-Distill also achieves better results. Unlike existing MLLM frameworks, which focus only on distilling the output logits, PL-Distill also targets the projector, effectively addressing dimension mismatches between the teacher's and student's audio embeddings. Experimental results demonstrate that the projector, as a key component for integrating different modal information, plays a crucial role in distillation.

\vspace{-0.1cm}
\subsection{Ablation Study}
\label{ssec:Ablation Study}
Table \ref{t2} presents an ablation study on the core components of the PL-Distill framework across three datasets, using UA and WA as the evaluation metrics. The results clearly demonstrate the significant effectiveness of both our proposed Projector-Level Distillation (PDist) and the AwCKA mechanism.
Specifically, by comparing LDist and LDist + PDist (CKA), it is observed that adding PDist (CKA) significantly improves the model’s performance across all metrics and datasets. This further underscores the critical role of Projector-Level Distillation in achieving effective modality alignment and demonstrates the effectiveness of the CKA method.
Further analysis, comparing LDist $+$ PDist (CKA) with LDist + PDist (AwCKA), underscores the significant impact of AwCKA. This supports our hypothesis that, in speech emotion recognition tasks, key emotional features are often concentrated in specific audio tokens. By leveraging attention scores from the LLM's final self-attention layer to dynamically weight these tokens, AwCKA effectively prioritizes these crucial time steps during CKA computation, consistently resulting in superior performance.
\vspace{-0.1cm}

\section{Conclusion}
\label{sec:Conclusion}
This paper presents the LALM knowledge distillation framework PL-Distill, applied to speech emotion recognition (SER), which successfully distilled an 8.4B parameter teacher model into a 1.1B parameter student model. Our approach focuses on Projector-Level Distillation (PDist) and introduces the Attention-weighted Centered Kernel Alignment (AwCKA), which enables the student model to better prioritize emotion-related audio tokens. Additionally, AwCKA addresses the challenge of dimension mismatches in the distillation process. Experimental results demonstrate that PL-Distill achieves excellent performance across multiple datasets. 
Beyond SER, PL-Distill has broader implications for audio-language modeling. By enabling effective knowledge transfer in LALMs (bridging acoustic and linguistic modalities), it facilitates efficient distillation for other audio tasks. AwCKA-driven preservation of task-relevant features also provides a general strategy for MLLM distillation requiring selective knowledge transfer. We will explore applying PL-Distill to more tasks in the future.

% \section{REFERENCES}
% \label{sec:refs}

% List and number all bibliographical references at the end of the
% paper. The references can be numbered in alphabetic order or in
% order of appearance in the document. When referring to them in
% the text, type the corresponding reference number in square
% brackets as shown at the end of this sentence \cite{C2}. An
% additional final page (the fifth page, in most cases) is
% allowed, but must contain only references to the prior
% literature.

% Please follow the IEEE Citation Guidelines, \url{https://ieee-dataport.org/sites/default/files/analysis/27/IEEE\%20Citation\%20Guidelines.pdf} for formatting of references.


% \vfill\pagebreak
\clearpage
\bibliographystyle{IEEEbib}
\bibliography{strings,refs}

\end{document}